\documentclass[%
reprint,
superscriptaddress,
 amsmath,amssymb,
aps,
prb,
]{revtex4-2}
\usepackage{soul}
\usepackage{pifont}
\usepackage{verbatim}
\usepackage{subcaption}
\usepackage{graphicx}%
\usepackage{dcolumn}%
\usepackage{bm}%
\usepackage{xcolor}
\usepackage[hidelinks]{hyperref}%
\usepackage{url}
\usepackage[normalem]{ulem} %
\usepackage[capitalize]{cleveref}
\usepackage{placeins}

\usepackage{physics}
\usepackage{amsthm,amsbsy,amsfonts,gensymb}
\usepackage{wrapfig}

\newcommand{\R}{\mathbb{R}}

\newcommand{\I}{\mathbb{I}}
\newcommand{\veps}{\varepsilon}

\usepackage[utf8]{inputenc}
\usepackage[T1]{fontenc}
\usepackage{pgfplots}
\usepgfplotslibrary{groupplots,dateplot}
\usetikzlibrary{patterns,shapes.arrows,math,arrows.meta}
\pgfplotsset{compat=newest}
\usepackage{tikz-3dplot}

\usetikzlibrary{positioning}

\newcolumntype{P}[1]{>{\centering\arraybackslash}p{#1}}
\newcolumntype{M}[1]{>{\centering\arraybackslash}m{#1}}

\usepackage{calligra}
\DeclareMathAlphabet{\mathcalligra}{T1}{calligra}{m}{n}
\DeclareFontShape{T1}{calligra}{m}{n}{<->s*[2.2]callig15}{}
\newcommand{\rdiff}{\ensuremath{\mathcalligra{r}}}

\begin{document}

\title{Glass Patterns in Twisted Disordered Crystals}

\author{Aaron Dunbrack}
\affiliation{Department of Physics and Nanoscience Center, University of Jyv\"askyl\"a, P.O. Box 35, FI-40014 University of Jyv\"askyl\"a, Finland}

\date{\today}

\begin{abstract}
    Twisting and stacking two copies of a 2D crystal can produce a long-wavelength periodic interference pattern known as a moir\'e pattern. Performing the same procedure with an aperiodic structure instead generates a single moir\'e spot at the rotation center, known as a \textit{Glass pattern}. We explore the implications of these patterns across a variety of models: they allow measurement of microscopic parameters from mesoscopic resistivity measurements; they generate an impurity that modifies the properties of a moir\'e lattice at the rotation center; and they allow for domain formation in amorphous magnets. These results establish Glass patterns as a generic feature of twisted disordered systems and provide a framework for future theoretical and experimental exploration.
\end{abstract}

\maketitle

\section{Introduction\label{Sec:Intro}}
Since the discovery of superconductivity and correlated insulator states in twisted bilayer graphene\cite{Cao_2018_CI,Cao_2018_SC}, there has been an explosion of research into multilayer materials with slightly mismatched lattice vectors, either due to nearly-identical lattice constants or a small relative twist between layers. Such systems are called \textit{moir\'e heterostructures} because the nearly-matched periodicity produces a beating phenomenon called a \textit{moir\'e pattern} \cite{amidrorVolOne}. This pattern can produce flat minibands, and the resulting increase in the density of states enhances physics driven by electron-electron interactions \cite{Bistritzer12233}.

Moir\'e effects are more general than nearly-identical lattices, however. Periodic moir\'e patterns can emerge if the lattices are instead nearly commensurate \cite{Rossi_2019,Dunbrack_2022,Scheer_2022,Dunbrack_2023,Crepel_2023,Scheer_2023}, or even without any approximate lattice matching \cite{amidrorVolOne,Dunbrack_2023,Putzer_2024}. In fact, one can produce moir\'e effects even in aperiodic situations \cite{amidrorVolTwo} --- for \textit{any} nontrivial configuration, one can observe local moir\'e effects by overlaying the configuration with a twisted copy of itself. This effect is called a \textit{Glass pattern} (named after their discoverer, Leon Glass \cite{glass1969moire,glass1976pattern,amidrorVolTwo}), and is illustrated in \cref{fig:glasspatterns}.

The key ingredient of these patterns is local cross-correlation between the disorder of the two layers \textit{before twisting}. This should be contrasted with independent disorder in the two crystals \cite{UCDis1,UCDis2}, or from disorder induced by the twist itself \cite{TADis1,TADis2,TADis3,TADis4,TADis5,TADis6}, neither of which produce Glass patterns. The necessary cross-correlations arise naturally if the two disordered layers are exfoliated from a single untwisted bilayer, rather than being obtained from independent pieces of material.

In what follows, we focus on the consequences of this correlated-disorder construction. \cref{Sec:twistedamorphous} provides an overview of Glass patterns in both amorphous and crystalline systems. \cref{Sec:MagGlass,Sec:adatomall,Sec:CorrImpur} then present various specific models of twisted disordered systems and examine the implications of the Glass pattern in these contexts.

\section{Theory of Glass Patterns}\label{Sec:twistedamorphous}
\begin{figure}
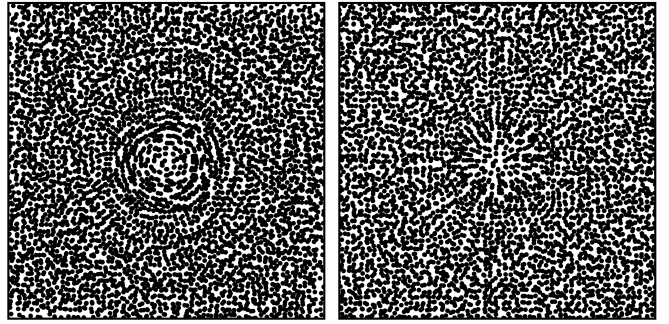

    \centering
    \include{Images_BasicGlassPatterns}
    \caption{Glass patterns arising from $5\degree$ twist (left) and $1.1\times$ relative scaling (right). For visual efficacy \cite{amidrorVolTwo}, the dots are located at randomly displaced lattice points rather than completely uncorrelated positions.}
    \label{fig:glasspatterns}
\end{figure}

We first illustrate Glass patterns using overlaid random-dot images, in the absence of crystalline order. Two effects arise near the deformation center, both of which can be seen in \cref{fig:glasspatterns}.

First, very near the center, the two layers have dots in the same place, resulting in an overall increased brightness compared to the uncorrelated dot locations away from the center. These \textit{intensity variations} are the most direct analogue of periodic moir\'e patterns, and are visible even at a distance \cite{amidrorVolOne,amidrorVolTwo,Dunbrack_2023}. The correlations that cause these intensity variations also underlie the physics discussed in \cref{Sec:CorrImpur,Sec:adatomall,Sec:MagGlass}.

Second, slightly off-center, the dots are not overlapping but the pairing remains visually apparent. The resulting visual structures are known as \textit{dot trajectories}, and in \cref{fig:glasspatterns} appear as concentric circles (left) and radial rays (right). They are only discernible at shorter distances, as one has to be able to resolve the individual dots to identify the local nematic order \cite{amidrorVolTwo}. The direction of these trajectories depends on the type of deformation. These finite-range correlations are relevant to the physics discussed in \cref{Sec:CorrImpur}.

In general, Glass patterns arise near any fixed point $x_0$ of a nonlinear transformation $x\rightarrow f(x)$, provided the derivative $Df(x_0)$ is approximately the identity matrix $\I$. The size of the Glass pattern is then inversely proportional to $||Df(x_0)-\I||$; see \cref{Apx:MathDetails} for details. For a rigid twist, $||Df-\I||\sim\theta$, so this is directly analogous to the corresponding inverse relationship between moir\'e pattern size and twist angle in the periodic case.

\begin{figure}
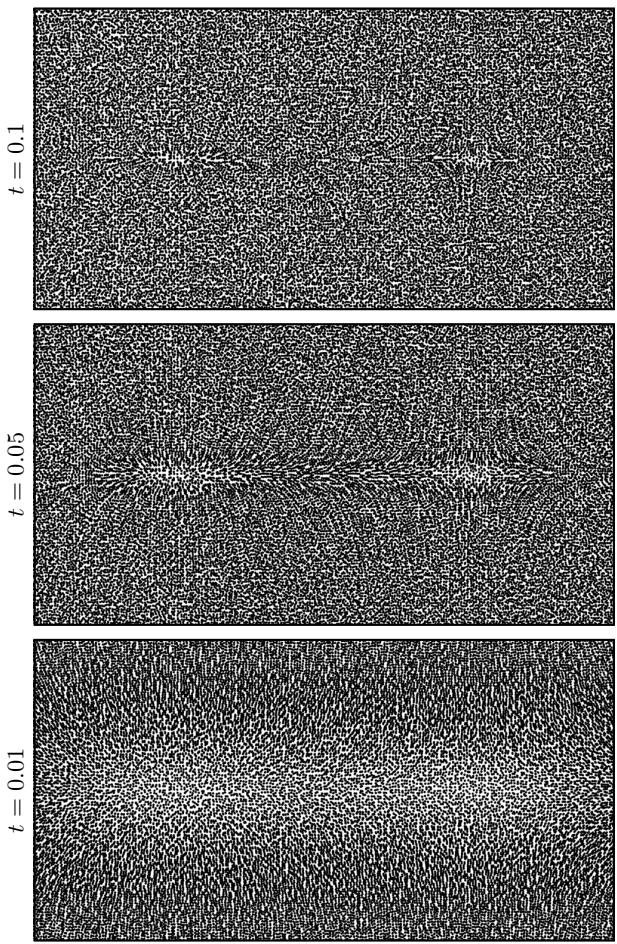

    \centering
    \include{Images_QuadraticGlassPatterns}
    \caption{Distorting one copy of an amorphous material by $(x,y)\rightarrow ((1+2t)x,y+t(y^2-1))$ results in two fixed points at $(0,\pm 1)$, each producing Glass pattern. Overlap increases with pattern size as $t$ decreases.}
    \label{fig:quadraticglass}
\end{figure}

If there are multiple fixed points, then one can have multiple Glass patterns. These Glass patterns can potentially overlap, depending on the details of the transformation, as illustrated in \cref{fig:quadraticglass}. Such tunable overlap is impossible for the repeating domains in a moir\'e lattice: the stackings in neighboring cells differ by a relative lattice translation, so the domain wall between them is unavoidable.

\begin{figure*}
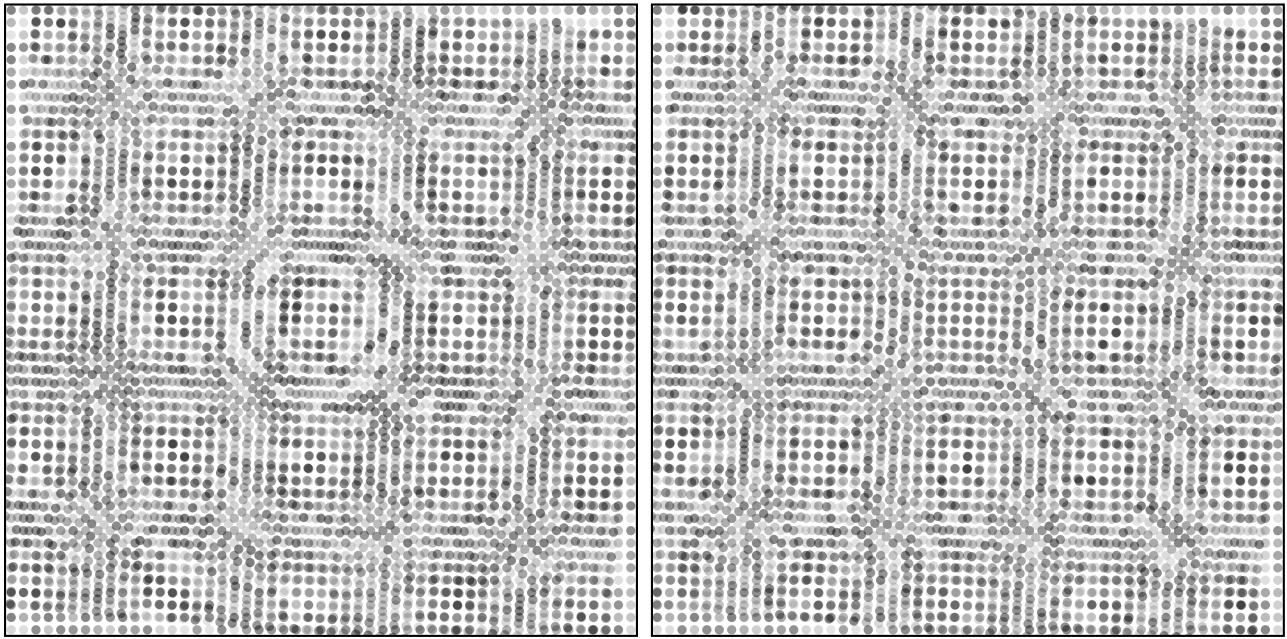

    \centering
    \include{Images_MoireGlass}
    \caption{Twisting two copies of the same disordered crystal produces both a moir\'e pattern (from the lattice) and a Glass pattern (from disorder correlations). Disorder is represented here by varying point opacities; the left figure is correlated, the right is anticorrelated. In the anticorrelated case, the central moir\'e cell shows enhanced uniformity, while in the correlated case dot trajectories persist across multiple cells. The relative prominence of these features reflects this visual construction rather than a universal property.}
    \label{fig:glassplusmoire}
\end{figure*}

When twisting crystals with cross-correlated disorder, Glass patterns are superimposed on the moir\'e pattern arising from crystal structure. An example of this combination of patterns is illustrated in \cref{fig:glassplusmoire}, wherein the disorder is represented as varying opacity of the individual dots. While the moir\'e pattern repeats itself on a lengthscale of $1/\theta$, the Glass pattern emerges only once, rendering the twist center identifiable as an effective impurity in the moir\'e lattice. Having an identifiable twist center (by contrast with clean moir\'e systems) is a consequence of breaking translation symmetry in the original microscopic lattice.

Under relative translation of the twisted layers, both the moir\'e and Glass patterns shift with the new rotation center, so the location of the Glass pattern \textit{within} the moir\'e unit cell is independent of the twist center. The glass pattern is typically colocated with the pre-twist stacking, but in general this depends on the nature of the microscopic correlations (see \cref{Apx:Glassloc}).

\section{Correlated Anderson disorder\label{Sec:CorrImpur}}

As our first model, we consider a twisted bilayer with interlayer-correlated Anderson disorder and compute the self-energy in the Born approximation (details in \cref{Apx:DisGrnDetails}). For simplicity, we consider only small-angle twist, approximated as $r\rightarrow (\I\pm\theta M/2)r$, where $\theta$ is twist angle and $M$ a $\pi/2$ rotation. Other transformations are analogous; e.g., for lattice mismatch, $\theta$ is the relative difference and $M=\I$.

The Green's function of the clean system is given by
\begin{equation}
    G^{(0)}=(E-H+i0^+)^{-1}.
\end{equation}
The dirty system is then described by the disorder-averaged Green's function. In the presence of some weak disorder $V(r)$, 
the disorder-averaged self-energy can be computed using the Born approximation
\begin{equation}
    \Sigma(r,r')\approx \langle V(r)G^{(0)}(r,r')V(r')\rangle.
\end{equation}

The form of disorder we consider is an on-site potential $V_i(r)$ in each layer $i$ with no intralayer correlations,
\begin{equation}
    \langle V_i(R+\rdiff/2)V_i(R-\rdiff/2)\rangle=\sigma^2\delta(\rdiff),
\end{equation}
but interlayer correlations diagonal in position before twisting,
\begin{equation}
    \langle V_1(R+\rdiff/2)V_2(R-\rdiff/2)\rangle=\alpha\sigma^2\delta(\rdiff-\theta MR),
\end{equation}
where $\alpha$ is a parameter determining the strength of these correlations (with $|\alpha|\leq 1$).

The intralayer component of self-energy is the same as the uncorrelated case,
\begin{equation}\label{eq:DiagSigmaGen}
    \Sigma_{ii}(R,k)=\sigma^2G^{(0)}_{ii}(R,\rdiff=0).
\end{equation}
However, there is an interlayer self-energy due to the correlations,
\begin{equation}\label{eq:OffDiagSigmaGen}
    \Sigma_{12}(R,k)=\alpha\sigma^2e^{-i\theta k\cdot MR}G^{(0)}_{12}(R,\rdiff=\theta MR).
\end{equation}
Note $\Sigma(R,k)$ is in the Wigner representation (its second argument is in momentum space), but $G^{(0)}(R,\rdiff)$ has its second argument in real space.
\cref{eq:DiagSigmaGen,eq:OffDiagSigmaGen} can also be understood as the self-consistency equations of the self-consistent Born approximation by replacing $G^{(0)}$ with the full Green's function $G$.

We now restrict to the simplest case: take two identical 2D electron gases with a uniform, isotropic interlayer hopping (in particular: no moir\'e modulation). Then $G^{(0)}(R,\rdiff)=G^{(0)}(\rdiff)$ is independent of $R$, yielding $\Sigma_{12}(R,k)=\alpha\sigma^2e^{-i\theta k\cdot MR}G^{(0)}_{12}(\theta MR)$ - i.e., up to the phase factor, the dependence of the self-energy on $R$ is the same as the microscopic Green's function's dependence on $\rdiff$, but enlarged to mesoscopic scales. With no underlying lattice, only the Glass pattern emerges.

The eigenvalues of the non-Hermitian part of the self-energy, which constitute the scattering rates of the particles, are
\begin{equation}\label{eq:scatteringrate}
    \tau_{\pm}^{-1}=-2\sigma^2(\Im[G_{11}^{(0)}(0)]\pm \alpha\Im[G_{12}^{(0)}(\theta MR)]).
\end{equation}
The $\pm$ signs correspond to the layer-symmetric and antisymmetric states at $R=0$.

If we simply treat the system as two parallel equally-populated channels $\pm$ and therefore approximate the resistivity as
\begin{equation}
    \rho\sim (\sum_\pm\tau_\pm)^{-1},
\end{equation}
then to leading order in $\alpha$ the change in resistivity is
\begin{equation}
    \frac{\delta\rho}{\rho}=-\alpha^2\left(\frac{\Im[G_{12}^{(0)}(\theta MR)]}{\Im[G_{11}^{(0)}(0)]}\right)^2,
\end{equation}
so the correlations reduce resistivity at all $R$. However, for $R\neq 0$ the symmetric and antisymmetric eigenstates of $G^{(0)}$ are not also eigenstates of $\Im(\Sigma)$, so the relationship between scattering time and resistivity is more complicated in the realistic mesoscopic transport problem.

We now consider limiting behavior. For large $R$, $G^{12}\rightarrow 0$, so \cref{eq:scatteringrate} reduces to the uncorrelated $\alpha=0$ case. In particular, the lifetimes of the two modes are equal,
\begin{equation}
    \tau_\pm^{-1}(R\rightarrow\infty)=\pi\sigma^2(\rho_++\rho_-),
\end{equation}
where $\rho_{\pm}(E)$ denotes the DOS of the dispersion $\veps_\pm$.

Next, we consider the behavior at $R=0$. In this case, the lifetime reduces to
\begin{equation}
    \tau_\pm^{-1}(R=0)=\pi\sigma^2[(1\pm\alpha)\rho_\pm+(1\mp\alpha)\rho_\mp].
\end{equation}
For perfectly correlated disorder with $\alpha=1$, this reduces to 
\begin{equation}
    \tau_\pm^{-1}(R=0)=2\pi\sigma^2\rho_\pm.
\end{equation}
Because near $R=0$ the disorder respects the vertical mirror symmetry $M_z$, the symmetric and antisymmetric sectors can be treated independently even at the level of disorder, and therefore their scattering rates are independent of each other.

For perfectly anticorrelated disorder with $\alpha=-1$ at $R=0$,
\begin{equation}
    \tau_\pm^{-1}(R=0)=2\pi\sigma^2\rho_\mp.
\end{equation}
Because the disorder scatters from one $M_z$ sector to the other, the scattering rate of each sector is proportional to the density of states of the \textit{opposite} sector. Thus, at any energies with bands from only one sector, the system behaves as if clean at $R=0$.

\begin{figure*}
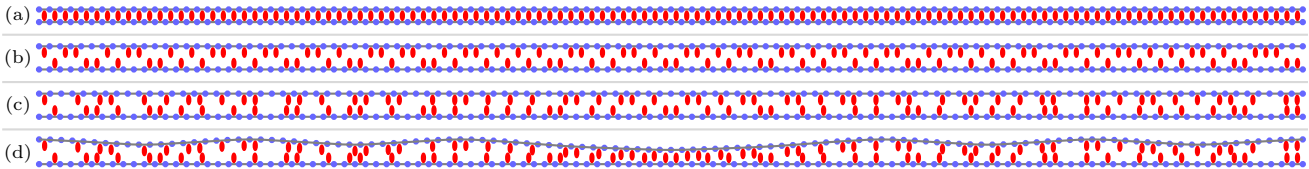

    \centering
    \include{Images_AdatomExample}
    \caption{Construction of the adatom model in \cref{Sec:adatomrelax}: (a) Begin with a bilayer crystal with interlayer adatoms. (b) Exfoliate such that the adatoms allocate randomly and independently between layers. (c) Twist or strain one layer. (d) Re-stack: anticorrelation at the center produces a Glass pattern in addition to the moir\'e modulation.}
    \label{fig:adatom_ex}
\end{figure*}

\begin{figure}
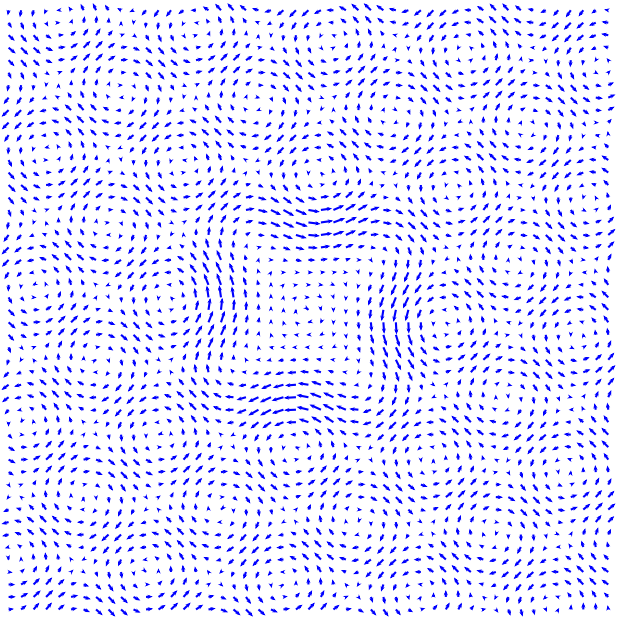

    \centering
    \include{Images_RelaxationVectorField}
    \caption{Using a 2D version of the model in \cref{fig:adatom_ex}, we fix the atomic locations in one layer and allow the other to relax in-plane. Here we indicate the displacement of that layer relative to a rigid twist. Note the changed behavior at the twist center: the Glass pattern inverts the preferred stacking from AB to AA, disrupting the periodic relaxation pattern.}
    \label{fig:adatomrelaxationvectorfield}
\end{figure}

For intermediate $R$, the local scattering time probes the bilayer's real-space Green's function on mesoscopic scales. For example: with interlayer hopping $t(r,r')=-t\delta(r-r')$ ($t>0$), the dispersion is $\epsilon_{\pm} = k^2/(2m) \mp t$ for mirror eigenvalues $\pm 1$. For energies between $-t$ and $t$, below the bottom of the $-$ band, the scattering times are
\begin{equation}
    \tau_{\pm}^{-1}=-\sigma^2(\Im[G_+(0)]\mp\alpha\Im[G_+(\theta MR)]).
\end{equation}
Measuring $\tau$ versus $R$ then maps the Green's function: $R \to \infty$ gives $\sigma^2 G_+(0)$, $R=0$ fixes $\alpha$, and intermediate $R$ yields $G_+(\theta MR)$ up to overall scaling.

The scattering time is measurable via local resistivity (extractable via, e.g., scanning tunneling potentiometry \cite{scanningtunnelingpotentiometry}), provided the mean free path is much less than the characteristic length of the Glass pattern, $v_F\tau\ll k_F^{-1}\theta^{-1}$. In conjunction with the Born approximation, this is equivalent to requiring a disorder strength in the range $E_F\theta\ll \hbar\tau^{-1}\ll E_F$. This probe remains useful for probing the bilayer Green's function even beyond the 2D electron gas, although the details differ due to the $R$-dependence of $G^{(0)}$.

\section{Adatom partitioning\label{Sec:adatomall}}
We next consider twisted disordered crystals in which the disorder originates from adatom configurations. In this system, both the Glass pattern from disorder and the moir\'e pattern from crystal structure contribute to the resulting physics.

The steps to construct this system are illustrated in \cref{fig:adatom_ex}. We begin with a bilayer with a regular interlayer adatom array, effectively a trilayer where central layer has weak intralayer bonds. When exfoliated, each adatom attaches to either the top or bottom layer, randomly and independently. This yields two disordered layers with perfectly anticorrelated adatom positions.

After twisting (or straining, as in \cref{fig:adatom_ex}), the top and bottom layers form a moir\'e pattern of AA (aligned) and AB (half-unit-cell shifted) stackings. At the AA-stacked twist center, the adatoms are anticorrelated; away from it, the adatom locations are functionally independent.

Some assumptions of this construction can be relaxed without changing the underlying physics involved. First, these results do not require the pre-exfoliation adatom pattern to be periodic, or in the specific locations shown in \cref{fig:adatom_ex}. The Glass pattern will arise for any configuration which is dense enough that the anticorrelation of adatom positions represent a significant constraint on the combined disorder. Second, the adatoms do not need to be fully independent on each site; it suffices that the domain size is much smaller than the moir\'e wavelength. Finally, the pre-twist stacking of the outer layers does not need to be AA; other stackings will change the location of the Glass pattern within the moir\'e unit cell.

\subsection{Lattice Relaxation\label{Sec:adatomrelax}}

First, we consider lattice relaxation effects, which are generally important in moir\'e systems \cite{LatticeRelaxReviewGr,LatticeRelaxReviewTMD}. As such, the adatoms will likely impact electron bandstructure even if they are electronically inert.

We model an interlayer attraction obstructed by adatoms, allowing out-of-plane relaxation, as illustrated in \cref{fig:adatom_ex}. In uncorrelated AA stacked regions, some adatoms stack vertically, doubling the original separation. In AB regions, the adatoms partially interleave, slightly reducing that spacing. Finally, at the twist center, the anticorrelation enables perfect meshing, restoreing the interlayer spacing of the original crystal.

If stacking energy is proportional to interlayer separation, then generic AA regions are energetically unfavorable compared to the AB regions, but the central AA region is energetically favorable. Accordingly, the Glass pattern reverses the usual moir\'e relaxation trend at the twist center. Allowing in-plane relaxation produces a set of deformations illustrated in \cref{fig:adatomrelaxationvectorfield} (details in \cref{Apx:simdetails}) which strongly distinguish the origin from equivalent moir\'e points.

\subsection{Local density of states in an adatom-based double-stub lattice\label{Sec:doublestub}}

We next analyze a 1D tight binding model for a particular such adatom construction. We construct couplings to highlight the impact of the Glass pattern on flat band physics. We compute the LDOS at zero energy and show that it has differing behavior at the center of the Glass pattern compared to other regions of the moir\'e lattice.

\begin{figure}
    \centering
    \begin{tikzpicture}[ultra thick, scale=1.0]

\draw (-3,-0.8) -- (3,-0.8);
\draw (-3,0.8) -- (3,0.8);

\foreach \x in {-3,-2,-1,0,1,2,3} {
    \draw[gray] (\x,-0.8) -- (\x,0.8);

    \fill (\x,0.8) circle [radius=0.15cm];
    \fill (\x,-0.8) circle [radius=0.15cm];
    \fill[red] (\x,0) circle [radius=0.1cm];

    \begin{scope}[green!60!black,<->,>={Stealth[scale=0.5]}]
        \draw (\x,0.8) ++ (0.1,-0.1) -- ++ (0,-0.6);
            \node at ({\x+0.25},0.45) {\tiny $\bm{t}'$};
        \draw (\x,-0.8) ++ (0.1,0.1) -- ++ (0,0.6);
            \node at ({\x+0.25},-0.35) {\tiny $\bm{t}'$};
        \ifnum\x<3
            \draw (\x,0) ++ (0.1,0) -- ++ (0.8,0);
                \node at ({\x+0.5},0.15) {\tiny $\bm{t}$};
        \fi
    \end{scope}
}

\begin{scope}[blue,>=latex]
    \draw[<->] (3.7,-0.8) -- (3.7,0.8);
    \node at (4,0) {$M_z$};
\end{scope}

\end{tikzpicture}
    \caption{Model in \cref{Sec:doublestub} before moir\'e effects: two atomic layers (black) with interlayer adatoms (red). Black lines indicate strong structural bonds, gray lines indicate weak ones. Green arrows denote electronic hoppings, which form a double-stub lattice. The $M_z$ symmetry of this construction from \cref{eq:Mz} is indicated in blue.}
    \label{fig:doublestubbase}
\end{figure}
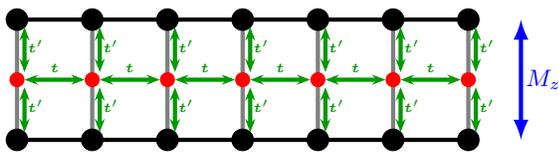

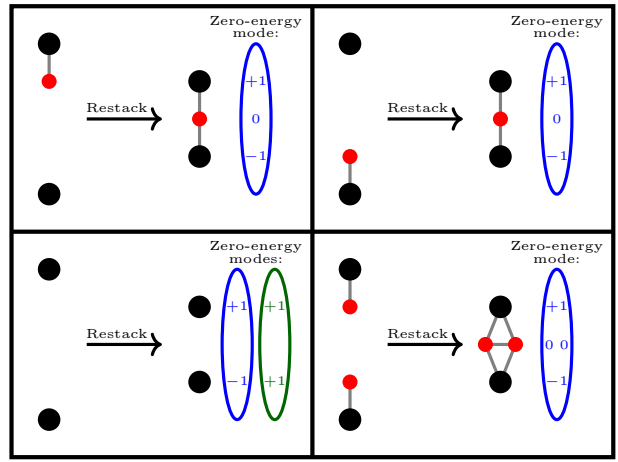
\begin{figure}
    \centering
    \begin{tikzpicture}[very thick]

\begin{scope}[ultra thick]
    \draw (4,0) -- (4,6);
    \draw (0,3) -- (8,3);
    \draw (0,0) rectangle (8,6);
\end{scope}

\newcommand{\elecorb}[3]{
    \begin{scope}[shift={(3.75,1.5)}]
        \draw (0,0) ellipse (0.2 and 1);
        \node at (0,0.5) {\tiny #1};
        \node at (0,0) {\tiny #2};
        \node at (0,-0.5) {\tiny #3};
    \end{scope}
}

\newcommand{\sectorsetup}[1]{
    \fill (1,2.5) circle [radius=0.15];
    \fill (1,0.5) circle [radius=0.15];
    \draw[->] (1.5,1.5) -- (2.5,1.5);
    \node at (1.9,1.65) {\tiny Restack};
    \fill (3,2) circle [radius=0.15];
    \fill (3,1) circle [radius=0.15];
    \node at (3.75,2.8) {\tiny Zero-energy};
    \node at (3.75,2.65) {\tiny mode{#1}:};
}

\newcommand{\centerlollypop}[3]{
    \draw[gray] (#1,#2) -- +(#3:0.5);
    \fill[red] (#1,#2) circle [radius=0.1];
}
\newcommand{\doublelollypop}[1]{
    \draw[gray] (#1,1.5) -- (3,2);
    \draw[gray] (#1,1.5) -- (3,1);
    \fill[red] (#1,1.5) circle [radius=0.1];
}

\begin{scope}[shift={(-0.5,0)}]

\begin{scope}
    \sectorsetup{s}
    \begin{scope}[blue,shift={(-0.25,0)}]
        \elecorb{$+1$}{}{$-1$}
    \end{scope}
    \begin{scope}[green!40!black,shift={(0.25,0)}]
        \elecorb{$+1$}{}{$+1$}
    \end{scope}
\end{scope}

\begin{scope}[shift={(4,0)}]
    \centerlollypop{1}{2}{90}
    \centerlollypop{1}{1}{-90}
    \draw[gray] (2.8,1.5) -- (3.2,1.5);
    \doublelollypop{2.8}
    \doublelollypop{3.2}
    \sectorsetup{}
    \begin{scope}[blue]
        \elecorb{$+1$}{$0\ 0$}{$-1$}
    \end{scope}
\end{scope}

\begin{scope}[shift={(0,3)}]
    \centerlollypop{1}{2}{90}
    \doublelollypop{3}
    \sectorsetup{}
    \begin{scope}[blue]
        \elecorb{$+1$}{$0$}{$-1$}
    \end{scope}
\end{scope}

\begin{scope}[shift={(4,3)}]
    \centerlollypop{1}{1}{-90}
    \doublelollypop{3}
    \sectorsetup{}
    \begin{scope}[blue]
        \elecorb{$+1$}{$0$}{$-1$}
    \end{scope}
\end{scope}
\end{scope}
\end{tikzpicture}
    \caption{After splitting and stretching the chains of Fig. 6, AA regions away from the origin yield four possible adatom configurations (panels). For three of four, only the $M_z=-1$ zero mode survives (blue). If neither layer contributes an adatom, an additional $M_z=+1$ zero mode appears (green).}
    \label{fig:doublestubstackings}
\end{figure}

For this model, we assume that the adatoms of the undeformed bilayer are at the same $x$-position as the atomic sites rather than offset by half a unit cell, as illustrated in \cref{fig:doublestubbase}. We then place a tight-binding model on this lattice with the same structural bonds as in the previous section (strong intralayer bonds in the top and bottom layers, no intralayer bonds in the central layer, and weak interlayer bonds; these are the black and gray lines in \cref{fig:doublestubbase}), but with a set of electronic hoppings not proportional to those bond strengths. We take the top and bottom layers as atomic insulators, but enable hopping from these layers to the adatoms with hopping $t'$. The central adatoms then form a 1D electronic chain with hopping $t$. This constructs a double-stub lattice (green arrows in \cref{fig:doublestubbase}) with effective Hamiltonian
\begin{equation}
    H=\begin{bmatrix}
        0&0&t'\I\\
        0&0&t'\I\\
        t'\I&t'\I&tH_c+2m\I
    \end{bmatrix}
\end{equation}
where $H_c$ is the matrix of hoppings for a 1D chain, with entries of $1$ on the superdiagonal and subdiagonal, and $2m$ is the chemical potential difference of the outer and inner layers.

This Hamiltonian features a mirror symmetry,
\begin{equation}\label{eq:Mz}
    M_z=\begin{bmatrix}0&\I&0\\\I&0&0\\0&0&\I\end{bmatrix}.
\end{equation}
Moreover, the set of $-1$ eigenvalues of $M_z$ are exact zero eigenvalues of $H$, producing a flat band at $E=0$. This flat band also follows from rank deficiency arguments analogous to those on bipartite lattices with a sublattice imbalance, corresponding to compact localized states with amplitude $\pm 1$ on the opposite-layer sites of each unit cell.

The other two bands occur at 
\begin{equation}
    E_{\pm}=m+t\cos(k)\pm\sqrt{(m+t\cos(k))^2+2t'^2},
\end{equation}
which for $t'\neq 0$ are gapped from zero energy.

We now suppose we separate the two layers as before, stretch one layer, and recombine them. However, in this case we assume the outer layers are rigid but the adatoms can freely twist in the central layer, such that after recombining the adatoms again bond symmetrically between the top and bottom layers; see \cref{fig:doublestubstackings}.

The effective Hamiltonian can now be expressed by some generic hopping matrices $T_i$ between the outer layers and the adatoms, and a hopping matrix $H_a$ between adatoms:
\begin{equation}
    H=\begin{bmatrix}0&0&T_1\\0&0&T_2\\T_1^\dagger&T_2^\dagger&H_a+2m\I\end{bmatrix}
\end{equation}

By the same rank-deficiency arguments, at least one third of the states remain at zero energy. The details of the construction will depend on the specific disorder realization and resulting matrix $T$, as well as the form of the adatom hopping $H_a$, but this lower bound on the number of states is independent of those details.

Near $AA$ regions, the Hamiltonian is still approximately $M_z$-symmetric, and moreover preserves the fact that the $M_z=-1$ states are zero energy (again following from the same argument as a bipartite construction). In $AB$ regions, the mirror symmetry is broken, and the construction becomes more complicated and disorder-dependent.

For concreteness and simplicity, we model $T$ as follows. Consider intersite spacing $a$; the hopping from a top or bottom layer atom at position $x$ to an adatom at position $\xi$ is $t'$ if $\xi<fa$ and $0$ otherwise, for $f$ some constant in the range $0.5<f<1$. I.e., we suppose the interlayer hopping is a step function with relative lateral displacement of the adatoms. The condition $f>0.5$ is set so that the AB regions have some nonzero hoppings between the outer layers and the adatoms.

\begin{figure}
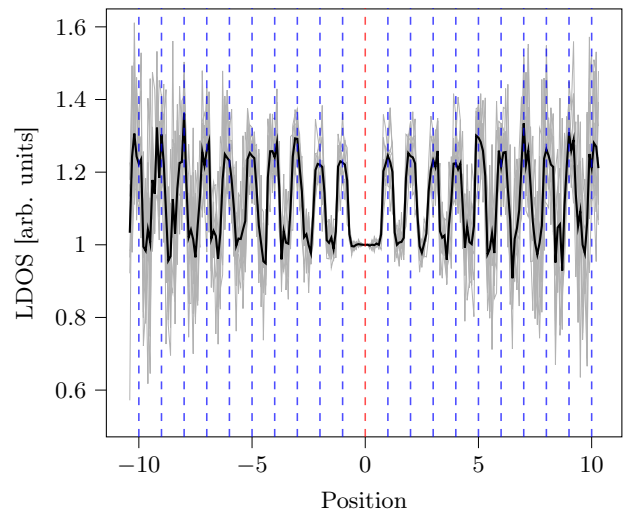

    \centering
    \include{Images_TightBinding075hard}
    \caption{Zero-energy LDOS of the model in \cref{Sec:doublestub} at mismatch $0.01$, $f=0.75$. Disorder realizations in gray, average in black. Vertical dashed lines mark the AA stacking points; the red line marks the strain center. Every AA region except the central one (due to the Glass pattern) experiences a 25\% LDOS enhancement.}
    \label{fig:doublestubLDOS}
\end{figure}

At the Glass pattern center, all zero-energy states have $M_z=-1$. In the other AA regions, some sites decouple from the adatoms, yielding zero-energy states with $M_z$ eigenvalue $+1$ that enhance the zero-energy LDOS by an average of 25\%. The fraction of the unit cell in which this enhancement occurs is determined by $f$. In \cref{fig:doublestubLDOS} we plot a numerical simulation of the same model, demonstrating this increase in the DOS. More numerical details and simulations with more complicated cutoffs are provided in \cref{Apx:simdetails}, but the qualitative behavior - periodic LDOS fluctuations that do not appear in the Glass pattern region - remains the same.

This model is particularly simple, as the flat bands are trivial --- the states' Wannier overlap vanishes, so they have no quantum geometry. However, the qualitative features should generalize to nontrivial flat-band models via the same mechanism: if the density of compact localized states differs between the clean and the dirty untwisted cases, then in the twisted case there will be a difference in LDOS between the twist center and the other equivalent points in the moir\'e pattern. For example, the same conclusions should hold in the diamond lattice, which has a similar isolated-atom construction.

\section{Magnetic domains in bilayer amorphous crystals\label{Sec:MagGlass}}
In addition to offering a way to design flat bands, moir\'e systems have also been applied to ferromagnetism. When interlayer magnetic coupling changes sign with stacking, sufficiently small twists form magnetic domains \cite{TwFM_Balents,TwFM_Review}. Intuitively, these domains form when the twist is small enough that the interlayer coupling  energy (proportional to domain area, $O(\theta^{-2})$) exceeds domain wall cost (proportional to domain perimeter, $O(\theta^{-1})$).

In twisted amorphous layers, the Glass pattern creates natural stacking domains: near the twist center the atoms are vertically aligned, whereas farther away their positions are independent. Thus, if interlayer coupling changes sign between these configurations, then a magnetic domain can form at the twist center.

\begin{figure}
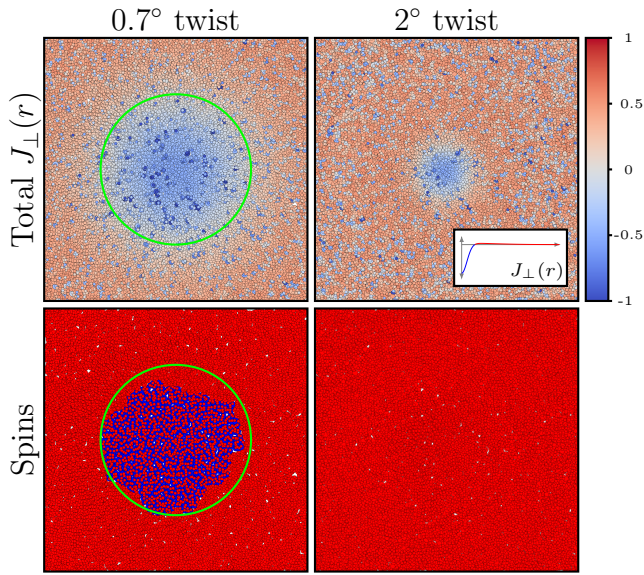

    \centering
    \include{Images_MagneticStates}
    \caption{Twisted bilayer amorphous Ising ferromagnet for two twist angles. Top: net interlayer coupling at each in-plane position, obtained by summing $J_\perp(r)$ (\cref{eq:Jperp}, plotted in inset) over all opposite-layer sites. $J_\perp(r)$ is antiferromagnetic for small $r$ but ferromagnetic for large $r$. Bottom: ground-state spin configurations, showing that at small twist the Glass pattern induces a central antiferromagnetic domain (see \cref{Apx:simdetails} for simulation details). Green circle indicates the theoretical domain radius from \cref{eq:magdomenbal}.}
    \label{fig:MagnetGlass}
\end{figure}

Concretely, let each layer be an identical copy of an amorphous ferromagnet with intralayer coupling $J$, and add interlayer magnetic coupling $J_\perp(x)$ which is ferromagnetic on average but antiferromagnetic for small displacements (or vice versa). For appropriate parameter choices, small twists produce a magnetic domain at the rotation center, as shown in \cref{fig:MagnetGlass} for intralayer $J=1$ and interlayer
\begin{equation}\label{eq:Jperp}
    J_\perp(r)=0.05\,e^{-(r/1.8)^2}-e^{-(r/0.3)^2},
\end{equation}
as plotted in the inset of \cref{fig:MagnetGlass}. In realistic systems, the relation is likely to be more complicated, arising from e.g. competition between a short-range direct exchange interaction and a long-range RKKY coupling.

We now derive the critical angle below which the magnetic domain appears for identical layers. For simplicity, we take $J_\perp(r)/J_\perp(0)$ to have a positive maximum at $r=0$, decrease to a single negative minimum, then increase to zero as $r\rightarrow\infty$, meaning $J(r)$ matches the qualitative features of \cref{eq:Jperp} up to a sign.

Let $\rho$ be the site density of each layer, $z$ the average intralayer coordination number, and $l$ the average bond length. Then, the average number of bonds within one layer cut per unit length by a random line is $C=zl\rho/\pi$. For uncorrelated layers, the net interlayer coupling per atom is $\bar J_\perp=\int_{\R^2} \rho J_\perp(r)d^2r$; we take $\bar J_\perp<J_\perp(0)$ to ensure that the correlations dominate at the center of the Glass pattern.

The domain radius $R^*$ for small $\theta$ is set by optimizing the domain energy with respect to $r^*=\theta R^*$,
\begin{multline}
    \frac{d}{dr}\Bigg[(2\pi r\theta^{-1})CJ+(\pi r^2\theta^{-2})\rho|\bar J_\perp|\\
    -\int_0^{r\theta^{-1}} 2\pi r'\rho|J_\perp(\theta r')|dr'\Bigg]_{r=r^*}=0.
\end{multline}
The first term is from the domain wall, the second from the average effect of $J_\perp$ between randomly-placed atoms, and the third the effect of vertical pairs. This then simplifies to
\begin{equation}\label{eq:magdomenbal}
    CJ\theta=\rho(|J_\perp(r^*)|-|\bar J_\perp|)r^*.
\end{equation}
Given the assumed properties of $J_\perp$, for sufficiently small $\theta$, this equation has two positive solutions $r^*$; only the larger solution provides an energy minimum. For large $\theta$, there are no solutions, indicating that such $\theta$ are above the critical angle. For the model and parameters used in \cref{fig:MagnetGlass} ($CJ\approx 4$, $\rho=1$, $\bar J_\perp\approx 0.23$), $R^*\approx 23$ at $0.7\degree$, and there is no solution at $2\degree$.

The existence of such a solution guarantees only a metastable solution; to identify the true ground state one must compare to the uniform $R^*=0$ state. This gives an equation for the critical angle,
\begin{equation}  
    \theta_c=\frac{\rho}{2CJr^*}\left[-r^{*2}|\bar J_\perp|+\int_0^{r^*} 2r'|J_\perp(r')|dr'\right],
\end{equation}
that must be solved in conjunction with the previous equation. The value of $r^*$ at the critical angle is dependent only on the profile of $J_\perp$,
\begin{equation}
    |J_\perp(r^*)|-|\bar J_\perp|/2=\int_0^1 s|J_\perp(r^*s)|ds,
\end{equation}
whereupon $\theta_c$ is most conveniently computed from \cref{eq:magdomenbal}. For the model and parameters used in \cref{fig:MagnetGlass}, $r^*\approx 0.26$ and $\theta_c\approx 0.84\degree$, meaning at minimum $R^*\approx 17.6$.

This estimate of $\theta_c$ is likely slightly conservative; the disorder could stabilize a domain wall beyond the mean-field threshold implied by a circular domain. However, in real samples, we expect that such fluctuation-induced shifts are less important to account for than hysteresis.

\section{Conclusion}
In this paper, we have explored various physical implications of Glass patterns, which arise when a twisted bilayer system has disorder with interlayer correlations that were local before twisting. These patterns can manifest in a variety of mesoscopic-local physical observables such as scattering time, lattice relaxation, local density of states, and magnetic domains. Notably, local resistivity measurements can offer a way to measure the interlayer component of the real-space Green's function on mesoscopic scales, as outlined in \cref{Sec:CorrImpur}.

We have largely set aside the issue of how to engineer such correlations, but offer brief speculation here. The prototypical method would be to begin with a bilayer system and exfoliate, either with the bilayer initially having layer-correlated disorder or with the disorder being generated by the exfoliation process (as in the adatom-based model of \cref{Sec:adatomall}). Another possibility would be to align the two layers with some space in between, then fire a single particle beam through both layers simultaneously to generate disorder; this would avoid requiring disorder-preserving exfoliation. Finally, one could grow two monolayers separately on the same disordered substrate as a way to engineer a somewhat consistent disorder pattern, provided the disordered substrate survives the exfoliation of the grown monolayer.

\section{Acknowledgments}
This work was supported by the Keele and Jane and Aatos Erkko foundations as part of the SuperC collaboration, by the Finnish Quantum Flagship (project no. 210000621611), and was inspired in part by the Nordita workshop ``Topology and Geometry Beyond Perfect Crystals.'' The author acknowledges useful conversations with Tim Kokkeler, Tero Heikkil\"a, and Lucas Baldo.

\bibliographystyle{apsrev4-2}
\bibliography{main}

\appendix
\section{Mathematical Formalism of Glass Patterns \label{Apx:MathDetails}}
We here provide a more precise formalism behind Glass patterns, broadly mirroring the explanations in Ref.~\cite{amidrorVolTwo}. The emergence of Glass patterns broadly result from three ingredients:
\begin{itemize}
    \item Before twisting, the two layers must have some correlation between them. We characterize this by a correlation function $C(r,r')$ which we assume to be small for $|r-r'|$ sufficiently large.
    \item The measurement instrument (for visual patterns, the human eye) must only be sensitive to sufficiently short-range correlations. We characterize the instrument sensitivity, as a function of $r-r'$, by $\chi(r-r')$. We also assume this is small for $|r-r'|$ sufficiently large.
    \item Finally, the two layers must be transformed via two different maps which agree at at least one point. For simplicity, we here assume only one layer is transformed via an invertible map $r\rightarrow f(r)$, and that $f$ has a single isolated fixed point at the origin, $f(0)=0$ (e.g., as occurs for twisting). $f$ should also satisfy some additional conditions described below (analogous to requiring small twist or mismatch for moir\'e patterns).
\end{itemize}

In this case, the Glass pattern emerges around the fixed point. The shape and structure of the Glass pattern depend on the Jacobian $J=Df(r=0)$ (as well as $C$ and $\chi$). Since the Glass pattern emerges for small $r$, we can make the approximation $f(r)\approx Jr$.

In periodic layers, periodic moir\'e pattern only arise from \textit{linear} transformations $f$ (up to moir\'e-periodic corrections from relaxation). I.e., one can have twisting or \textit{spatially uniform} strains, but arbitrary deformations will yield aperiodic moir\'e patterns. Glass patterns do not have such global constraints because they are local behavior near fixed points, and so they are much more generic.

Long-wavelength moir\'e patterns arise only for \textit{small} twist angles where $Df\approx\I$. This condition also applies to Glass patterns, where we also demand $J\approx\I$ (more precisely $||J-\I||\ll 1$). We also make the additional nondegeneracy constraint that $\det(J-\I)\neq 0$ so as to get a 2D rather than 1D Glass pattern, again by analogy with moir\'e patterns \cite{amidrorVolOne,Dunbrack_2023}.

To clarify why a single disordered configuration admits a correlation function $C(r,r')$, we define it explicitly for the dot-image case. Each image $i$ is represented by a transparency function $T_i(r)$ such that the composite image has transparency $T(r)=T_1(r)T_2(r)$ (see Refs \cite{amidrorVolOne,amidrorVolTwo,Dunbrack_2023}). The correlation of the two layers $C(r,r')$ can be understood as a local spatial average
\begin{equation}
    C(r,r')=\int_{B_R} T_1(r+\epsilon)T_2(r'+\epsilon)d\epsilon
\end{equation}
for $B_R$ a ball of radius $R$ centered at $\epsilon=0$. $R$ should be in the range $a\ll R\ll ||J-I||^{-1}a$ to average over local disorder without washing out spatial variations of the original images at the Glass pattern scale. In the spatially-uniform case $C(r,r')=C(r-r')$, the upper limit is not required, and so
\begin{equation}\label{eq:imagecorrfn}
    C(r,r')\rightarrow C(\delta)=\int_{\R^2} T_1(r)T_2(r+\delta)dr.
\end{equation}

From this point forward, we take the simplifying assumption that $C(r,r')=\delta(r-r')$. A realistic $C$ resulting from overlapping dots would have some width proportional to the radius of the dots that we neglect. In fact in \cref{fig:glasspatterns}, because the dots are not completely random and are instead displaced from lattice sites, the correlation function is actually negative in some region when $r-r'$ is larger than dot radius but less than the lattice constant; this is the origin of the darker region immediately around the bright center in that figure.

\subsection{Intensity variations}
The intensity variations are characterized by higher-than-usual (local) correlations - i.e., that the dots of the two layers are aligned. The relevant observable, therefore, is the correlation function itself, weighted by the cutoff,
\begin{equation}
    O_{IV}(r)=\int C(r,f^{-1}(r'))\chi(r-r')dr'
\end{equation}
which in the case of images with $C(r,r')$ as in \cref{eq:imagecorrfn} and $\chi$ a delta function quantifies the local brightness from 0 to 1.

We now take $C(r,r')=\delta(r-r')$ and the approximation $f(r)=Jr$, with $J\approx \I$ so $\det(J)\approx 1$ (to leading order; this is exact for twist). We can then evaluate the integral over $r'$ as simply
\begin{equation}
    O_{IV}(r)=\chi((\I-J)r).
\end{equation}
Therefore, under the assumption that we are only sensitive to local correlations (i.e., that $\chi$ is sharply peaked at $0$), $O_{IV}$ has a broad peak near the origin: the original peak of $\chi$, transformed by the matrix $(\I-J)^{-1}$ (which is large in magnitude since $I\approx J$).

\subsection{Dot trajectories}
We now consider the dot trajectories that usually surround the central region. These are instead characterized by the extent to which dots tend to make pairs displaced by a small amount, and therefore have a vector (or nematic) order parameter characterizing the displacement,
\begin{equation}\label{eq:DTvector}
    O_{DT}(r)=\int (r-r')C(r,f^{-1}(r'))\chi(r-r')dr'
\end{equation}
By taking the same approximations as the previous argument, this integral evaluates to
\begin{equation}
    O_{DT}(r)=\chi((\I-J)r)(\I-J)r.
\end{equation}

In this case, the domain of the visual effect is more toroidal: for large $r$ it is suppressed by $\chi$ as with the central region, but for small $r$ it is suppressed by the linear term. The direction of these trajectories is determined by the particular transformation $f$ via the Jacobian $J$ at the fixed point. In the case of twist, the trajectories are circular, whereas for isotropic strain, the trajectories are radial, as illustrated in \cref{fig:glasspatterns}.

We note that the cutoff function $\chi$ for dot trajectories is not necessarily the same as the one for the central region. Indeed, for dot images the characteristic decay length of $\chi$ in the central region is the radius of the dot (determining the region where dots overlap), whereas the length for dot trajectories is on the order of the typical space between dots (over which length the eye can still pick out the pairs).

Note the order parameters of the central region and dot trajectories differ in their dimensions. Assuming $\chi$ is a dimensionless function with a peak magnitude of order $1$ and a cutoff of some microscopic length scale $a$, then $O_\text{CR}$ will similarly be dimensionless of order $1$ near its peak. Meanwhile, $O_\text{DT}$ will instead be of order $a$, a microscopic length. For this reason, the central region is more easily observed at large distances than the dot trajectories. The observables that can be resolved at long distances constitute what is called the \textit{macrostructure} of these patterns, whereas the remaining effects (including dot trajectories) constitute the \textit{microstructure} \cite{amidrorVolTwo}.

\section{Glass pattern location\label{Apx:Glassloc}}
In this appendix, we demonstrate that what matters isn't just fixed point of the deformation, but also the structure of the correlation function $C(x,x')$ defined in \cref{Apx:MathDetails}. 

The simplest correlation function would start with AA stacking and be perfectly correlated at corresponding sites and uncorrelated elsewhere, such that $C(x,x')=\delta(x-x')$. This would center the Glass pattern on the strain-center AA stacking.

The analogous construction for AB stacking has a slightly nonlocal the correlation function. In a 1D chain, a maximally-symmetric correlation function would take the form $C(x,x')=k[\delta(x-x'+a/2)+\delta(x-x'-a/2)]$. This produces a Glass pattern centered on the strain-center AB stacking.

However, the AB correlations can arise in AA stacking with the correlation function $C(x,x')=k[\delta(x-x')+\delta(x-x'+a)]$. Then one finds a Glass pattern split across two AA regions, one part at the strain center and another shifted by a moir\'e unit cell. This is equivalent to a pattern centered at the AB region in between, demonstrating that what matters is the AB-stacked correlation and not the actual AA stacking.

The AB correlation function in an AA-stacked chain is rather artificially constructed, breaking natural symmetries. However, in crystals more complicated than a 1D chain where the initial stacking is not highly symmetric, this can arise naturally. There is also not necessarily a clean quantification of location in these more complicated scenarios, as there may be multiple types of correlations with different Glass pattern centers.

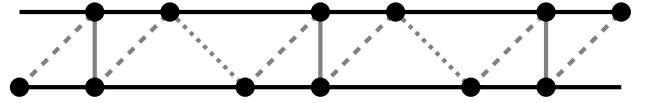
\begin{figure}
    \centering
    \begin{tikzpicture}[ultra thick]

    \draw (0,0) -- (8,0);
    \draw (0,1) -- (8,1);
    \begin{scope}[gray]
    \begin{scope}[]
        \draw (1,0) -- (1,1);
        \draw (4,0) -- (4,1);
        \draw (7,0) -- (7,1);
    \end{scope}
    \begin{scope}[dashed]
        \draw (0,0) -- (1,1);
        \draw (1,0) -- (2,1);
        \draw (3,0) -- (4,1);
        \draw (4,0) -- (5,1);
        \draw (6,0) -- (7,1);
        \draw (7,0) -- (8,1);
    \end{scope}
    \begin{scope}[dotted]
        \draw (2,1) -- (3,0);
        \draw (5,1) -- (6,0);
    \end{scope}
    \end{scope}
    
    \filldraw (0,0) circle (0.1cm);
    \filldraw (1,0) circle (0.1cm);
    \filldraw (1,1) circle (0.1cm);
    \filldraw (2,1) circle (0.1cm);
    \filldraw (3,0) circle (0.1cm);
    \filldraw (4,0) circle (0.1cm);
    \filldraw (4,1) circle (0.1cm);
    \filldraw (5,1) circle (0.1cm);
    \filldraw (6,0) circle (0.1cm);
    \filldraw (7,0) circle (0.1cm);
    \filldraw (7,1) circle (0.1cm);
    \filldraw (8,1) circle (0.1cm);
\end{tikzpicture}
    \caption{Bilayer system with layers in black, interlayer correlations in gray. Upon straining one layer, the Glass pattern from the solid-line correlations would be centered in an AB region, from the dashed lines in an AA region, and from the dotted lines in a BA region, demonstrating that what matters is not the initial stacking, but the details of the correlations.}
    \label{fig:BernalCorrStrip}
\end{figure}

Consider a 1D strip of Bernal bilayer graphene along the armchair direction, as illustrated in \cref{fig:BernalCorrStrip}.
Suppose there is both an on-site AB coupling and a nearest-neighbor AA/BB coupling. The former correlation would produce a Glass pattern centered at the strain center AB region, but the latter would be centered at the AA region adjacent.

\section{Computation of disorder-averaged Green's functions\label{Apx:DisGrnDetails}}
In this section, we provide more algebraic detail behind the results in \cref{Sec:CorrImpur}. We begin with a Green's function $G^{(0)}_{ij}(R,\rdiff)$ of the clean twisted bilayer, expressed in terms of center-of-mass and relative coordinates, and assume an on-site disorder potential $V(r)$ with correlation function
\begin{equation}
    C_{ij}(R-\rdiff/2,R+\rdiff/2)=\sigma^2[f(\rdiff)\delta_{ij}+\alpha h(\rdiff-\theta MR)\tau^x_{ij}].
\end{equation}
The parameter $\theta$ and matrix $M$ encode the relative transformation $T$ (e.g., for a twist, $T=R(\theta)$); $f$ and $h$ set the range of intra- and interlayer correlations, respectively, before the twist. In the main text, $f$ and $h$ are both assumed to be delta functions.

In the Wigner basis,
\begin{align}
    \bar C_{ij}(R,k)=&\int e^{-ik\cdot\rdiff}C_{ij}(R+\rdiff/2,R-\rdiff/2)d\rdiff\nonumber\\
    =&\sigma^2[\hat f(k)\delta_{ij}+e^{-i\theta k\cdot MR}\hat h(k)\tau^x_{ij}]
\end{align}
and
\begin{equation}
    \bar G_{ij}^{(0)}(R,k)=\int \frac{d^2k}{(2\pi)^2}e^{-ik\cdot\rdiff}G_{ij}^{(0)}(R+\rdiff/2,R-\rdiff/2)
\end{equation}

The Born approximation correction is then
\begin{equation}
    \Sigma_{ij}(R,k)=[\bar C_{ij}*\bar G_{ij}](R,k)
\end{equation}
where $*$ indicates convolution in the second index.

For delta function correlations $\hat f=\hat h=1$, the convolution simplifies to
\begin{align}
    \Sigma_{ii}(R,k)&=\sigma^2 \int\frac{d^2q}{(2\pi)^2}\bar G^{(0)}_{ii}(R,q)=\sigma^2 G^{(0)}_{ii}(R,0)\\
    \Sigma_{12}(R,k)&=\alpha\sigma^2\int\frac{d^2q}{(2\pi)^2}e^{-i\theta (k-q)\cdot MR}\bar G^{(0)}_{12}(R,q)\nonumber\\
    &=\alpha\sigma^2 e^{-i\theta k\cdot MR}G^{(0)}_{12}(R,\theta MR)
\end{align}

If we take the two layers to be identical and the interlayer hopping to have continuous rotational and translational symmetry, then the two layers of the clean system are equivalent even after twist, so we can write $G_{11}^{(0)}=G_{22}^{(0)}=g$ and $G_{12}^{(0)}=G_{21}^{(0)}=g'$. Since $g$, $g'$ are even and layer interchange corresponds to $\epsilon\rightarrow -\epsilon$, the clean Green's function can be decomposed into its layer-symmetric and antisymmetric sectors,
\begin{equation}
    G_{\pm}^{(0)}=g\pm g'=(E-\veps_{\pm}+i0^+)^{-1},
\end{equation}
where $\veps_\pm(k)$ are the energies of the symmetric and antisymmetric states. Inversely,
\begin{equation}
    g=(G^{(0)}_++G^{(0)}_-)/2,\qquad g'=(G^{(0)}_+-G^{(0)}_-)/2.
\end{equation}

In this same decomposition, the components of self-energy are
\begin{equation}\begin{split}
    \Sigma_{\pm\pm}(R,k)=&\sigma^2\left[g(0)\pm\alpha \cos(\theta k\cdot MR)g'(\theta MR)\right],\\
    \Sigma_{\pm\mp}(R,k)=&\pm i\alpha\sigma^2 \sin(\theta k\cdot MR)g'(\theta MR).
\end{split}\end{equation}
The eigenvalues of this matrix are
\begin{equation}
    \tau_{\pm}^{-1}=-2\sigma^2(\Im[g(0)]\pm \alpha\Im[g'(\theta MR)]),
\end{equation}
where the $\pm$ signs correspond to layer-symmetric and antisymmetric states at $R=0$.

In the special case of delta-function hopping, $H_\text{interlayer}=t(c_{r,1}^\dagger c_{r,2}+h.c.)$, between two identical 2d electron gases, the dispersion is $\epsilon_{\pm}(k)=\epsilon_0(k)\pm t$, where $\epsilon_0(k)=\frac{k^2}{2m}$ is the 2d electron gas dispersion.

Suppose we consider states below the bottom of the upper band, so only $G_-$ is nonzero. Then,
\begin{equation}
    \tau_{\pm}^{-1}=-\sigma^2(\Im[G_-(0)]\mp\alpha\Im[G_-(\theta MR)]).
\end{equation}
Hence, if we determine the value of $\sigma^2$ by the large-$R$ limit and the value of $\alpha$ by the $R=0$ limit, then the full real-space Green's function can be mapped out by considering intermediate values of $R$.

\section{Simulation details\label{Apx:simdetails}}
In this appendix we summarize the more technical aspects of the code used to generate the figures shown in the main text for the sake of replicability, as well as address what happens with variants of the model in \cref{Sec:doublestub} with more realistic interlayer hoppings.

\subsection{Lattice relaxation figures}
We first consider the model of \cref{Sec:adatomrelax} and \cref{fig:adatom_ex,fig:adatomrelaxationvectorfield} therein. The functional forms are illustrative and chosen for qualitative demonstration.

The top layer in the bottom panel of \cref{fig:adatom_ex} is the sum of a cosine and an exponential, designed to roughly form-fit the adatoms drawn in between each other. The exact function, with mismatch $\delta=0.05$, is
$$y=0.25\cos(2\pi x\delta)-\exp(-8x^2\delta^2)+1.$$

For \cref{fig:adatomrelaxationvectorfield}, we suppose the bottom layer is fixed and model the displacement of the top layer as a continuous strain field. We take isotropic Lam\`e parameters for the top layer $\mu=\lambda=10$ and take a twist angle of 10 degrees. We then add a potential energy from the other layer, in terms of the local relative displacement of layers $u$ and moir\'e length $L_m$, of
\begin{equation}
    E[u]=0.5(\cos(2\pi u_x/L_m) + \cos(2\pi u_y/L_m))-3\exp(-|u|^2/8)
\end{equation}
with the first term representing the instability of AA stacking in general via the moir\'e potential and the second term the stability of AA stacking at zero displacement from the Glass pattern. The initial $u$ is chosen according to the twist, then allowed to relax to minimize the above energy.

\subsection{Tight binding model}
We now consider the model of \cref{Sec:doublestub}. For the parameters, we use $\delta=0.01$ mismatch using 2600 unit cells on each layer with open boundary conditions (only inner 2000 unit cells shown). The inter-site hopping between adatoms is taken to be exponentially small in separation, decay constant one unit cell, cut off at $2.5$ unit cells, with maximum hopping the same as the interlayer hopping.

For the LDOS, Lorenzian broadening is used over energy with width $10^{-3}$, and sites in real-space are weighted into the LDOS using a Gaussian with a width of $5$ unit cells (cut off at $20$ unit cells). Results are averaged over 10 disorder realizations. Each unit of position is one moir\'e unit cell, i.e., $1/\delta$ atomic unit cells.

\begin{figure}
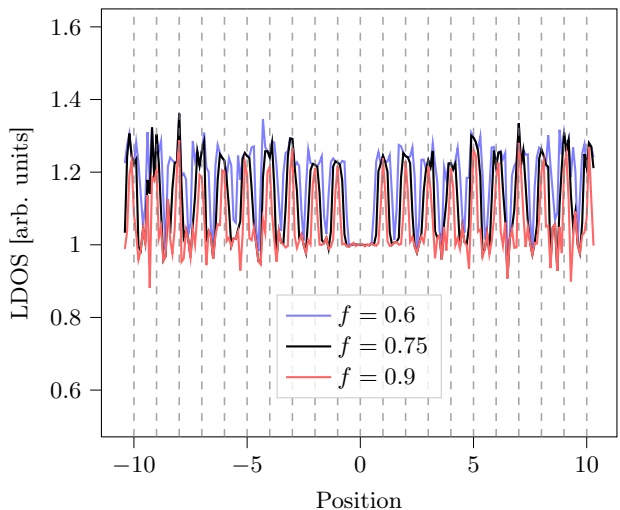

    \centering
    \include{Images_TightBindingAllHard}
    \caption{Zero-energy LDOS for the model in \cref{Sec:doublestub} with the same hard cutoff as shown in \cref{fig:doublestubLDOS}. Peaks get narrower with decreasing $f$.}
    \label{fig:TBAllHard}
\end{figure}

Varying the value of $f$ with the same disorder realizations produces the plot shown in \cref{fig:TBAllHard}. As can be seen, the height of the LDOS peaks remains approximately the same, but the width decreases with increasing $f$. This is because increasing $f$ reduces the fraction of space which is AA-like, in which adatoms bond to only one pair of sites each.

\begin{figure}
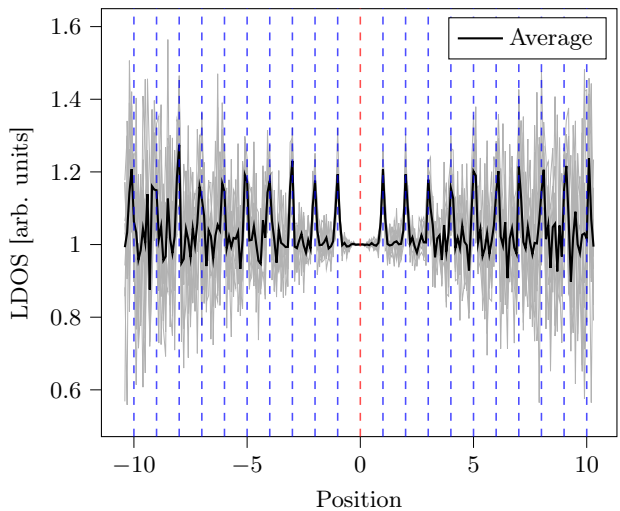

    \centering
    \include{Images_TightBindingSoft}
    \caption{Zero-energy LDOS for binding model from \cref{Sec:doublestub} with exponential interlayer hopping suppression instead of sharp cutoffs. Peaks from \cref{fig:doublestubLDOS} persist, showing the Glass pattern is robust.}
    \label{fig:TBSoft}
\end{figure}

A more realistic model, in which the interlayer hoppings are also cut off exponentially (decay constant $0.4$ unit cells, cut off at $2.5$ unit cells) is shown in \cref{fig:TBSoft}. The peaks have different behavior but still appear at the same points, showing that the Glass pattern is generic across hopping profiles.

\subsection{Magnetic structure}
To generate \cref{fig:MagnetGlass}, we construct a disordered system by taking a unit square lattice for each layer and adding random displacements, distributed as a Gaussian with width $0.3$, before twisting by the relevant angle. For each site, we then add a constant ferromagnetic coupling $J=1$ to its six nearest neighbors (producing a total $J=2$ if the nearest-neighbor relation holds in both directions). We then add an interlayer coupling $J_\perp(r)=0.05*e^{-(r/1.8)^2}-e^{-(r/0.3)^2}$ to all interlayer sites within $5$ units, which is antiferromagnetic for small $r$ but ferromagnetic for large $r$.

To avoid boundary artifacts, the four outer rings of each layer are pinned up. One spin at the origin in one layer is also pinned up to pick a unique ground state. This ground state is then computed with a Metropolis-Houdayer algorithm. Updates include single-site and vertical-pair flips, allowing domains to move coherently across layers.

We perform 500 Houdayer cycles \cite{houdayer2001cluster} with 100 Metropolis cycles each, on a linear schedule from $T=4$ to $T=0.01$, for each angle, after which we compare energy to a ferromagnetic state (to avoid finding domains that are only metastable); the results for the ground state in \cref{fig:MagnetGlass} are the result of that process. The figures are drawn with dot size indicating layer, although the individual dots are only visible upon very close examination.

\end{document}